\documentclass{aa}
\usepackage{graphics}
\usepackage{epsfig}
\usepackage{txfonts}
\def\beqa{\begin{eqnarray}}
\def\eeqa{\end{eqnarray}}
\def\beq{\begin{equation}}
\def\eeq{\end{equation}}

\def\half{\frac{1}{2}}

\def\gd{g_{\mu\nu}}

\def\gam{\gamma}
\def\alp{\alpha}

\def\sig{\sigma}

\def\dmunu{_{\mu\nu}}

\def\da{_{\alpha}}

\def\beqa{\begin{eqnarray}}
\def\eeqa{\end{eqnarray}}
\def\beq{\begin{equation}}
\def\eeq{\end{equation}}

\def\half{\frac{1}{2}}

\def\gd{g_{\mu\nu}}

\def\dmunu{_{\mu\nu}}

\def\da{_{\alpha}}

\let\gam=\gamma
\let\alp=\alpha
\let\sig=\sigma

\let\lb=\label
\renewcommand{\epsilon}{\varepsilon}

\def\p{\phi}

\def\p{\phi}

\let\gam=\gamma
\let\alp=\alpha
\let\sig=\sigma

\begin{document}

\title{Constraining scalar-tensor quintessence by cosmic clocks }
\author{S.\,Capozziello \inst{1} \fnmsep \inst{2}, P.K.S. Dunsby
\inst{3},E.\,Piedipalumbo\inst{1}\fnmsep \inst{2},
C.\,Rubano\inst{1}\fnmsep \inst{2} } \offprints{S. Capozziello,
capozziello@na.infn.it} \institute{Dipartimento di Scienze Fisiche,
Universit\`{a} di Napoli ``Federico II'' \and Istituto Nazionale di
Fisica Nucleare, Sez. Napoli, Via Cinthia, Compl. Univ. Monte S.
Angelo, 80126 Naples, Italy \and Department of Mathematics and
Applied Mathematics, University of Cape Town and \\South African
Astronomical Observatory, Observatory Cape Town, South Africa.}
\titlerunning{Constraining scalar-tensor quintessence by cosmic clocks}
\authorrunning{S. Capozziello \& al.}
\date{Received / Accepted}
\abstract{}{To study scalar tensor theories of gravity with power
law scalar field potentials as cosmological models for accelerating universe, using cosmic clocks.}{Scalar-tensor
quintessence models can be constrained by identifying suitable
cosmic clocks which allow to select confidence regions
for cosmological parameters. In particular, we constrain the
characterizing parameters of  non-minimally coupled scalar-tensor
cosmological models which admit exact solutions of the Einstein
field equations. Lookback time to galaxy clusters at low
intermediate, and high redshifts is considered. The high redshift
time-scale problem is also discussed in order to select other cosmic clocks such as
quasars. }{The presented models seem to work in all the regimes considered:
the main feature of this approach is the fact that cosmic clocks are
completely independent of each other, so that, in principle, it is
possible to avoid biases due to primary, secondary and so on
indicators in the cosmic distance ladder.  In fact, we have used different methods to test the
models at low, intermediate and high redshift by different
indicators: this seems to confirm independently the proposed dark
energy models.}{} \maketitle

\section{Introduction}

%%%%%%%%%%%%%%%%%%%%%%%%%%%%%%%%%%%%%%%%%%%%%%%%%%%%%%%%%%
An increasing harvest of observational data seems indicate that
$\simeq 70\%$ of the present day energy density of the universe is
dominated by a mysterious "dark energy" component, described in the
simplest way using the well known cosmological constant $\Lambda$
%explanation is the well known  cosmological constant $\Lambda$
(\cite{Perl97,perlal,rei&al98,Riess00}) and explains the accelerated
expansion of the observed Universe, firstly deduced by luminosity distance
measurements.
%and should solve the shortcomings of standard
%Einstein--Friedman cosmology by the inflationary paradigm.
However, even though the presence of a dark energy component is
appealing in order to fit observational results with theoretical
predictions, its fundamental nature still remains a completely open question.

Although several models describing the dark energy component have
been proposed in the past few years, one of the first physical
realizations of quintessence was a cosmological scalar field, which
dynamically induces a repulsive gravitational force, causing an
accelerated expansion of the Universe.

The existence of such a large proportion of dark energy in the
universe presents a large number of theoretical problems. Firstly,
why do we observe the universe at exactly the time in its history
when the vacuum energy dominates over the matter (this is known as
the \textit{cosmic coincidence} problem). The second issue, which
can be thought of as a \textit{fine tuning problem}, arises from the
fact that if the vacuum energy is constant, like in the pure
cosmological constant scenario, then at the beginning of the
radiation era the energy density of the scalar field should have
been vanishingly small with respect to the radiation and matter
component. This poses the problem, that in order to explain the
inflationary behaviour of the early universe and the late time dark
energy dominated regime, the vacuum energy should evolve and cannot
simply be \textit{constant}.

A recent work has demonstrated that the fine-tuning problem can be
alleviated by selecting a subclass of quintessence models, which
admit a \textit{tracking behaviour} (\cite{stein2}), and in fact, to
a large extent, the study of scalar field quintessence cosmology is
often limited to such a subset of solutions. In scalar field
quintessence, the existence condition for a tracker solution    provides
a sort of selection rule for the potential $V(\phi)$ (see
(\cite{all03}) for a critical treatment of this question), which
should somehow arise from a high energy physics mechanism (the so
called \textit{model building problem}). Also adopting a phenomenological
point of view, where the functional form of the potential $V(\phi)$
can be determined from observational cosmological functions, for
example the luminosity distance, we still cannot avoid a number of
problems. For example,  an attempt to reconstruct the potential from
observational data (and also fitting the existing data with a linear
equation of state) shows that a violation of the weak energy
condition (WEC) is not completely excluded (\cite{cald}), and this
would imply a \textit{superquintessence regime}, during which
$w_{\phi}<-1$ (\textit{phantom regime}). However it turns out that,
assuming a dark energy component with
an arbitrary scalar field Lagrangian, the transition from regimes with $w_{\phi}%
\geq-1$ to those with $w_{\phi}<-1$ (i.e. crossing the so called
\textit{phantom divide}) could be physically impossible since they
are either described by a discrete set of trajectories in the phase
space or are unstable (\cite{hu,vik}). These shortcomings have been
recently overcome by considering the \textit{unified phantom
cosmology} (\cite{odintsov}) which, by taking into account a
generalised scalar field kinetic sector, allows to achieve
models with natural transitions between inflation, dark matter, and
dark energy regimes. Moreover, in recent works, a dark energy
component has been modeled also in the framework of scalar tensor
theories of gravity, also called extended quintessence (see for
instance \cite{star,fuji2,francesca1,DGP1,chiba, amendola,
uzan,curvature,curvature2,MetricRn,PalRn1,PalRn2,Allemandi}).
Actually it turns out that they are compatible with a ${\it
peculiar}$ equation of state $w\leq -1$, and provide a possible link to the issues of
non-Newtonian gravity (\cite{fuji2}). In such theoretical background
the accelerated expansion of the universe results in an
observational effect of a non-standard gravitational action. In
extended quintessence cosmologies (EQ) the scalar field is coupled
to the Ricci scalar, $R$, in the Lagragian density  of the theory:
the standard term  $ \displaystyle 16 \pi G_{\ast}\,\,R$ is replaced
by $\displaystyle 16 \pi F(\phi)\,\, R$, where $F(\phi)$ is a
function of the scalar field, and $G_{\ast}$ is the {\it bare}
gravitational constant, generally different to the Newtonian
constant $G_N$ measured in Cavendish-type experiments (\cite{star}).
Of course, the coupling is not arbitrary, but it is subjected to
several constraints, mainly arising from the time variation of the
constants of nature (\cite{uzan2}). In EQ models, a scalar field has
indeed a double role: it determines at any time the effective
gravitational constant and contributes to the dark energy density,
allowing some different features with respect to the minimally
coupled case (\cite{uzan2}). Actually, while in the framework of the
minimally coupled theory we have to deal with a fully relativistic
component, which becomes homogeneous on scales smaller than the
horizon, so that standard quintessence cannot cluster on such
scales, in the context of nonminimally coupled quintessence theories
the situation changes, and the scalar field density perturbations
behave like the perturbations of the dominant component at any time,
demonstrating in the so called {\it gravitational dragging}
(\cite{francesca1}).

In this work, we focus our attention on the effect that dark energy
has on the background evolution (through the $t(z)$ relation) in the
framework of some scalar tensor quintessence models, for which exact
solutions of the field equations are known. In particular we show
that an accurate determination of the age of the universe together
with new age determinations of cosmic clocks can be used to produce
new strict constraints on these dark energy models, by constructing
the time-redshift $(t-z)$ relation, and comparing the theoretical
predictions with the observational data.

Although discrepancies between age determinations have long plagued
cosmology, the situation has changed dramatically in recent years:
type Ia supernova measurements, the acoustic peaks in the CMB
anisotropies (\cite{spergel}), and so on, are all
consistent with an age $t_{0}\simeq14\pm1Gyr$. Recently Krauss \&
Chaboyer (\cite{KC03}) have provided constraints on the equation of
state of the dark energy by using new globular cluster age
estimates.

Furthermore, Jimenez \& al. proposed a non parametric measurements
of the time dependence of $w(z),$ based on the relative ages of
stellar populations (\cite{Jimenez1,Jimenez2}). It is therefore
rather timely to investigate the implications of new age
measurements within in the framework of quintessence cosmology.

It turns out that such $(t-z)$ relations are strongly varying
functions of the equation of state $w(z),$ and could also be very
useful for breaking the degeneracies that arise in other
observational tests.

As a first step toward this goal, we study the cosmological
implications arising from the existence of the quasar APM 08279+5255
and three extremely red radio galaxies at $z=1.175$ (3C65), $z=1.55$
(53W091) and $z=1.43$ (53W069) with a minimal stellar age of 4.0
Gyr, 3.0 Gyr and 4.0 Gyr, respectively, extending the analysis
performed in (\cite{lima00,al03}) to a scalar-tensor field
quintessential model (\cite{cqg,nmc,ruggiero}).

As a further step, we also consider the \textit{lookback time} to
distant objects, which is observationally estimated as the
difference between the present day age of the universe and the age
of a given object at redshift $z,$ already used in (\cite{look}) to
constrain dark energy models. Such an estimate is possible if the
object is a galaxy observed in more than one photometric band, since
its color is determined by its age as a consequence of stellar
evolution.

It is thus possible to get an estimate of the galaxy age by
measuring its magnitude in different bands and then using stellar
evolutionary codes. It is worth noting, however, that the estimate
of the age of a single galaxy may be affected by systematic errors
which are difficult to control.

It turns out that this problem can be overcome by considering a
sample of galaxies belonging to the same cluster. In this way, by
averaging the age estimates of all the galaxies, one obtains an
estimate of the cluster age and thereby reducing the systematic
errors. Such a method was first proposed by Dalal \& al.
(\cite{DAJM01}) and then used by Ferreras \& al. (\cite{FMT03}) to
test a class of models where a scalar field is coupled to the matter
term, giving rise to a particular quintessence scheme. We improve
here this analysis by using a different cluster sample
(\cite{andreon,andreon2}) and testing a scalar tensor quintessence
model. Moreover, we add a further constraint to better test the dark
energy models and assume that the age of the universe for each model
is in agreement with recent estimates. Note that this is not
equivalent to the lookback time method as we will discuss below.

The layout of the paper is as follows. In Sect.\thinspace 2, we
briefly present the class of cosmological models which we are
going to consider, defining also the main quantities which we need
for the lookback time test. Sects.\thinspace 3 and \thinspace 4
are devoted to the discussion of cosmic clocks at low,
intermediate and high redshifts whose observational data are used
to test the theoretical background model. Finally we summarize and
draw conclusions in Sect.\thinspace V. The lookback time method is
outlined in the Appendix A, referring to (\cite{look}) for a
detailed exposition.
%%%%%%%%%%%%%%%%%%%%%%%%%%%%%%%%%%%%%%%%%%
\section{The model}
%%%%%%%%%%%%%%%%%%%%%%%%%%%%%%%%%%%%%%%%%%
The action for a scalar-tensor theory, where a generic
quintessence scalar field is non-minimally coupled with gravity,
and a minimal coupling between matter and the quintessence field
is assumed, is:
\begin{equation}
\mathcal{A}=\int_{T}\sqrt{-g}\left(
F(\phi)R+\frac{1}{2}g^{\mu\nu}\phi_{,\mu
}\phi_{,\nu}-V(\phi)+\mathcal{L}_{m}\right)  \,, \label{e1}%
\end{equation}
where $F(\phi),$ and $V(\phi)$ are two generic functions,
representing the coupling with geometry and the potential
respectively, $R$ is the curvature scalar,
${\displaystyle\frac{1}{2}g^{\mu\nu}\phi_{,\mu}\phi_{,\nu}}$ is
the kinetic energy of the quintessence field $\phi$ and
$\mathcal{L}_{m}$ describes the standard matter content. In units
$8\pi G= \hbar =c=1$ and signature (+,\,-,\,-,\,-),  we recover
the standard gravity for $F=-1/2$, while the effective
gravitational coupling is ${\displaystyle G_{eff}=-\frac{1}{2F}}$.
Chosing a spatially flat Friedman-Robertson-Walker metric in Eq. (\ref{e1}), it is possible to obtain the  \emph{pointlike} Lagrangian
\begin{equation}
\mathcal{L}=6Fa\dot{a}^{2}+6F^{\prime}\dot{\phi}a^{2}\dot{a}+a^{3}p_{\phi}-D\,,
\label{e2}%
\end{equation}
where $a$ is the expansion parameter and $D$ is the initial
dust-matter density. Here prime denotes derivative with respect to
$\phi$ and dot the derivative with respect to cosmic time. The dynamical
equations, derived from Eq. (\ref{e2}), are
\begin{equation}
H^{2}=-\frac{1}{2F}\left(
\frac{\rho_{\phi}}{3}+\frac{\rho_{m}}{3}\right)  \,,
\label{e3}%
\end{equation}%
\begin{equation}
2\dot{H}+3H^{2}=\frac{1}{2F}p_{\phi}\,, \label{e4}%
\end{equation}
where the pressure and the energy density of the $\phi$-field are
given by
\begin{equation}
p_{\phi}=\frac{1}{2}\dot{\phi}^{2}-V(\phi)-2(\ddot{F}+2H\dot{F})\,, \label{q-pressure}%
\end{equation}%
\begin{equation}
\rho_{\phi}=\frac{1}{2}\dot{\phi}^{2}+V(\phi)+6H\dot{F}\,, \label{q-density}%
\end{equation}
and $\rho_{m}$ is the standard matter-energy density. Considering
Eq. (\ref{e2}), the variation with respect to $\phi$ gives the
Klein-Gordon equation
\begin{equation}
\ddot{\phi}+3H\dot{\phi}+12H^{2}F^{\prime}+6\dot{H}F^{\prime}+V^{\prime}=0\,.
\label{klein-gordon}%
\end{equation}
%%%%%%%%%%%%%%%%%%%%%%%%%%%%%%%%%%%
\subsubsection{The case of power law potentials}
%%%%%%%%%%%%%%%%%%%%%%%%%%%%%%%%%%%
Recently it has been shown that it is possible to determine the
structure of a scalar tensor theory, without choosing any specific
theory a priori, but instead reconstructing the scalar field
potential and the functional form of the scalar-gravity coupling
from two observable cosmological functions: the luminosity distance
and the linear density perturbation in the dustlike matter component
as functions of redshift. Actually the most part of works where
scalar-tensor theories were considered as a model for a variable
$\Lambda$-term followed either a \textit{reconstruction} point of
view, either a traditional approach, with some special choices of
the scalar field potential and the coupling. However it is possible
also a \textit{third road} to determine the structure of a scalar
tensor theory, requesting some general and physically based
properties, from which it is possible to select the functional form
of the coupling and the potential. An instance of such a procedure
has been proposed in  (\cite{ruggiero2,ruggiero}): it turns out that
requiring the existence of a  Noether symmetry for the action in Eq.
(\ref{e1}), it is possible not only to select several analytical
forms both for $F(\phi)$ and $V(\phi)$, but also obtain exact
solutions for the dynamical system (\ref{e3},\ref{e4},\ref{klein-gordon}).
In this section we analyze a wide class of theories derived from such a
Noether symmetry approach, which show power law couplings and
potentials, and admit a {\it tracker behaviour}. Let us summarize the basic
results, referring to (\cite{nmc,nmc2}) for a detailed
exposition of the method. It turns out that the Noether symmetry exists only when
\begin{equation}
V = V_0 (F(\phi))^{p(s)} \,, \label{e5}
\end{equation}
where $V_0$ is a constant and
\begin{equation}\label{ps}
p(s)= \frac{3 (s+1)}{2s +3}\,,
\end{equation}
where $s$ labels  the class of such Lagrangians which admit a Noether
symmetry. A possible choice of $F(\phi)$ is
\begin{equation}
F = \xi (s) (\phi+\phi_{i})^2 \,, \label{e6}
\end{equation}
where
\begin{equation} {\displaystyle \xi(s)= \frac{(2 s +3)^2}{48 (s+1) (s+2)}}\,, \label{xi1}
\end{equation}
 and $\phi_{i}$ is an integration constant. The general solution
corresponding to such a potential and coupling is:
\begin{equation}
\label{general2}
 a(t)=A(s)\left(B(s)t^{3\over
 {s+3}}+{D\over {\Sigma_{0}}}\right)^{{s+1}\over
 s}t^{{2s^{2}+6s+3}\over {s(s+3)}}\,,
\end{equation}
\begin{equation}
 \label{generalphi2}
\phi(t)=C(s)\left(-{V_{0}\over \gamma(s)}B(s)t^{3\over
{s+3}}+{D\over \Sigma_{0}}\right)^{-{{2s+3}\over
2s}}t^{-{(2s+3)^{2}\over {2s(s+3)}}}\,,
\end{equation}
where  $A(s)$, $B(s)$, $C(s)$, $\gamma(s)$ and $\chi(s)$ are given
by
\begin{eqnarray}
  A(s)& =&\left( {\chi(s)}\right)^{s+1\over s}  \left({(s+3) \Sigma_0 \over 3\gamma(s) }\right)^{s+2\over
  s+3}\,,\\
  B(s) &=& \left({(s+3) \Sigma_0 \over 3\gamma(s) }\right)^{-{3\over (s+3)}}{(s+3)^2 \over s+6}\,,\\
  C(s) &= &\left({\chi(s)}\right)^{-{(2s+3)\over 2 s}}\left({(s+3) \Sigma_0 \over 3\gamma(s) }\right)^{-{(3+2s)\over
  2(s+3)}}\,,
\end{eqnarray}
and
\begin{eqnarray}
  \gamma(s) &=& {2 s+3\over 12 (s+1) (s+2)}\,,\\
  \chi(s)&=& -{  2 s\over 2 s+3}\,,
\end{eqnarray}
where $D$ is the matter density constant, $\Sigma_{0}$ is an integration constant
resulting from the Noether symmetry, and $V_{0}$ is
the constant which determines the scale of the potential. Even if
these constants are not directly measurable, they can be rewritten
in terms of cosmological observables like $H_0$, $\Omega_m$ etc, as in the
following:
\begin{eqnarray}
  D &=& \left(\left(\frac{1}{A(s)}\right)
   ^{\frac{s}{s+1}}-B(s)\right)
   \Sigma_0 \,,\\
  \Sigma_0 &=& \left(3^{-\frac{5
   s+6}{s^2+4 s+3}}
   (s+3)^{-\frac{3 s^2+7
   s+3}{s^2+4 s+3}} (s+6)\right.\times \nonumber\\
   &&\times\left.\frac{\left((\widehat{H}_0-2) s^2+3 (\widehat{H}_0-2)
   s-3\right) \gamma
   (s)^{\frac{s^2-s-3}{s^2+4
   s+3}}}{(s+1) \chi
   (s)}\right)^{\frac{(s+1)
   (s+3)}{s^2-s-3}}.
\end{eqnarray}
Here we are following the procedure used in (\cite{nmc}), taking the
age of the universe, $t_0$, as a unit of time. Because of our choice
of time unit, the expansion rate $H(t)$ is dimensionless, so that
our Hubble constant is not (numerically) the same as the $H_0$ that
appears in the standard FRW model and measured in ${\rm km s^{-1}
Mpc^{-1}}$. Setting $ a_0= a(t_0)=1$ and $\widehat{H}_0=H(t_0)$, we
are able to write $\Sigma_0$ and $D$ as functions of $s$ and
$\widehat{H}_0$. Since  the effective gravitational coupling is
$G_{eff}=-{1\over {2F}}$, it turns out that in order to recover an
attractive gravity we get $s \in (-2\, , \, \, -1)$. Restricting
furthermore the values of $s$ to the range $s \in (-{3\over 2}\, ,
\, \, -1)$ the potential $V (\phi)$ is an inverse power-law,
$\phi^{-2|p(s)|}$.\, In this framework, we obtain naturally an
effective cosmological constant:
\begin{equation}
\Lambda_{eff}=G_{eff}\rho_{\phi}, \label{lambdaeff}%
\end{equation}
where
$\lim_{t\rightarrow\infty}{\Lambda_{eff}}=\Lambda_{\infty}\neq0$. It
is then possible to associate an \textit{effective density
parameter} to the $\Lambda$ term, via the usual relation
\begin{equation}
\Omega_{\Lambda_{eff}}\equiv{\frac{\Lambda_{eff}}{3H^{2}}}. \label{omegaeff}%
\end{equation}
Eqs. (\ref{general2}) and (\ref{generalphi2})  are all that is
needed to perform the $t-z$ analysis. It is worth noting that, since
the lookback time - redshift $(t-z)$ relation does not depend on the
actual value of $t_{0}$, it furnishes an independent cosmological
test through age measurements, especially when it is applied to old
objects at high redshifts. For varying $w$, as in the case of our
model, the lookback time can be rewritten in a more general form by
considering
\[
\mathcal{H}_{0}t_{0}=\int_{0}^{1}{\frac{H_{0}da}{aH(a)}},
\]
and then writing
\[
t_{L}(z)=t_{H}\int_{0}^{\bar{a}}{\frac{H_{0}da}{aH(a)}},
\]
where $\bar{a}=1/(1+z)$ and $t_{H}={\frac{1}{H_{0}}}$ is the
\textit{Hubble time}.

\subsubsection{The case of quartic potentials}
%%%%%%%%%%%%%%%%%%%%%%%%%%%%%%%%%%%
As it is clear from Eq.(\ref{general2}) and
Eq.(\ref{generalphi2}) for a generic value of $s$ both the scale
factor $a(t)$ and the scalar field $\phi(t)$ have a power law
dependence on time. It is also clear that there are some additional
particular values of $s$, namely $s=0$ and $s=-3$ which should be
treated independently. Actually $s=0$ reduces to the minimally
coupled case (see \cite{pv}) while $s=-3$ corresponds to the induced
gravity with a  quartic potentials: namely,
$F=\frac{3}{32}\phi^{2}$, and $V(\phi)=\lambda \phi^{4}$ . This case
is particularly relevant since it allows one to recover the
self-interaction potential term of several finite temperature field
theories. In fact, the quartic form of potential is required in
order to implement the symmetry restoration in several Grand Unified
Theories. Consequently, we limit our analysis to this model, showing
that it naturally provides an accelerated expansion of the universe
and other interesting features of dark energy models.\,\,It worth to
note that since the \textit{special} derivation of our model, we can
focus the analysis on quantities concerning, mostly, the background
evolution of the universe, without recurring to any observable
related to the evolution of perturbations. The following analysis
can be easily extended to more general classes of non-minimally
coupled theories, where exact solutions can be achieved by Noether
symmetries (\cite{ruggiero,pla}), but, for the sake of simplicity,
we restrict only to the above relevant case. The general solution is
\begin{eqnarray}
&  a(t)=a_{i}e^{{\frac{-\alpha_{1}t}{3}}}\left[  {\left(  -1+e^{\alpha_{1}%
\,t}\right)  \,+\alpha_{2}\,t+\alpha_{3}}\right]  ^{\frac{2}{3}},\\
&  \phi=\phi_{i}\sqrt{\frac{e^{-\alpha_{1}t}}{\left(  -1+e^{\alpha_{1}%
\,t}\right)  \,+\alpha_{2}\,t+\alpha_{3}}}, \label{eqaphi}%
\end{eqnarray}
where $\alpha_{1}=4\sqrt{\lambda}$, and $a_{i}$ , $\alpha_{2}$
$\alpha_{3}$ and $\phi_{i}$ are integration constants, related to
the initial matter density, $D$, by the relation
$\alpha_{1}\alpha_{2}=D$, which implies that they cannot be null.
The case $\lambda=0$ has to be treated separately.

It turns out that $\alpha_{3}$, $\phi_{i}$ and $a_{i}$ have an
immediate physical interpretation: $a_{i}$ is connected to the value
of the scale factor at $t=t_{0}=1$, while $\alpha_{3}$ set the value
$a(0)$. We can

safely set $a(0)=0$, so that $\alpha_{3}=0$. Actually this is not
strictly correct, as our model does not extend up to the initial
singularity. However, this position introduces a shift in the time
scale which is small with respect to that of the radiation dominated
era. This alters by a negligible amount the value of $t_{0}$, while,
moreover, it is left undetermined in our parametrisation.

Since \thinspace\ \thinspace$F(\phi)\propto\phi^{2}$, and $G_{eff}%
(\phi)\propto-{\frac{1}{F(\phi)}}$, we note that a way of recovering
attractive gravity is to chose $\phi_{i}$ as a pure imaginary
number. Without compromising the general nature of the problem, we
can set $\phi_{i}=\imath$, so that our field in Eq. (\ref{eqaphi})
becomes  a purely imaginary field, giving rise to an apparent
inconsistency with the choice in the Lagrangian  (\ref{e2}).
Actually,  the generic infinitesimal generator of the Noether
symmetry is
\begin{equation}
X=\alpha \frac{\partial }{\partial {a}}+\beta \frac{\partial
 }{\partial {\phi }}+\dot{\alpha}\frac{\partial }{\partial
 {\dot{a}}}+\dot{\beta}\frac{\partial }{\partial {\dot{\phi}}},
\end{equation}
where $\alpha $ and $\beta $ are both  functions of $a$ and $\phi $,
and:
\begin{equation} \dot{\alpha}\equiv\frac{\partial
\alpha }{\partial
 a}\dot{a}+\frac{\partial
\alpha }{\partial \phi }\dot{\phi}\quad ;\quad
 \dot{\beta}\equiv\frac{\partial
\beta }{\partial a}\dot{a}+\frac{\partial \beta }{\partial \phi
 }\dot{\phi}.
\end{equation}
 Demanding the existence of Noether symmetry
$\mathcal{L}_{X}L=0$,
 we get
the following equations,
\begin{equation}
\alpha +2a\frac{\partial {\alpha }}{\partial
{a}}+a^{2}\frac{\partial
 {\beta}}{\partial {a}}\frac{F^{\prime }}{F}+a\beta \frac{F^{\prime }}{F}=0
\label{eq:i}
\end{equation}
\begin{equation}
\label{eq:iii} \left( 2\alpha +a\frac{\partial {\alpha }}{\partial
 {a}}+a\frac{\partial
\beta }{\partial \phi }\right) F^{\prime }+aF^{\prime \prime }\beta
 +2F\frac{\partial \alpha }{\partial \phi }+\frac{a^{2}}{6 }\frac{\partial{\beta}}{\partial {a}}=0
\end{equation}
\begin{equation}
3 \alpha +12 F'(\phi )\frac{\partial {\alpha }}{\partial
 {\phi}}+2 a \frac{\partial {\beta }}{\partial
 {\phi}}=0 \label{eq:ii}
\end{equation}
\begin{eqnarray}
&&{V^{\prime } \over V} = p(s){F^{\prime } \over F} \label{eq:iv}
\end{eqnarray}
It turns out that these equations are preserved if we adopt the more
common signature $(-, +,+,+)$ in the metric tensor and flip the sign
of the kinetic term.  This imply that the generic infinitesimal
generator of the Noether symmetry is preserved, and hence provides
 \textit{phantom} solutions of the field equations. Therefore, as far as
the \textit{quaestio} of setting $\phi_i=\imath$ concerns, it
turns out that the scalar field can be physically interpreted as a
 \textit{phantom} solution of the field equations adopting the
 signature $(-, +,+,+)$, even if it is mathematically represented
as a pure imaginary field.  It worth noting that  such phantom
solutions in scalar tensor gravity do not produce any violation in
energy conditions, since $\rho_{\phi}$, $\rho_{\phi} + p_{\phi}$
and $\rho_{\phi} +3 p_{\phi}$  are strictly positive in the minimally
coupled case (see \cite{Ellis} and references therein for a
discussion). Furthermore, complex scalar fields (in particular
purely imaginary ones) have been widely considered in classical
and quantum cosmology with minimal and non-minimal couplings
giving rise to interesting boundary conditions for inflationary
behaviors \cite{Khalatnikov1,Khalatnikov2}.

%%%%%%need a citation

Finally, assuming $H(1)=\mathcal{H}_{0}$, we get
\begin{eqnarray}
&  a(t)=a_{i}e^{-\frac{\alpha_{1}t}{3}}\left(
e^{\alpha_{1}t}-\frac{\left( 3\mathcal{H}_{0}\left(
e^{\alpha_{1}}-1\right)  -\alpha_{1}\left( e^{\alpha_{1}}+1\right)
\right)  t}{3\mathcal{H}_{0}+\alpha_{1}-2}-1\right)
^{2/3}\label{scale}\\
& \nonumber\\
&  \phi^{2}(t)=-\frac{e^{\alpha_{1}t}}{e^{\alpha_{1}t}-\frac{\left(
3\mathcal{H}_{0}\left(  e^{\alpha_{1}}-1\right)  -\alpha_{1}\left(
e^{\alpha_{1}}+1\right)  \right)
t}{3\mathcal{H}_{0}+\alpha_{1}-2}-1},
\label{fi}%
\end{eqnarray}
where%
\begin{equation}
a_{i}=e^{\frac{\alpha_{1}}{3}}\left(  \frac{3\mathcal{H}_{0}+\alpha_{1}%
-2}{2+2e^{\alpha_{1}}(\alpha_{1}-1)}\right)  ^{2/3}.
\end{equation}

We have to note that ${\displaystyle\delta\equiv{\frac{\dot{G}_{eff}}{G_{eff}}}
={\frac{3\mathcal{H}_{0}-\alpha_{1}}{2}}}$, and still $\alpha_{1}
\simeq 3\mathcal{H}_{0}$, as we shall see in the next section, this
expression is very small, as indicated by observations. This fact means that all physical processes implying the effective gravitational constant are not dramatically affected in the framework of our model.
%%%%%%%%%%%%%%%%%%%%%%%%%%%%%%%%%%%%%%%%%%%%%%%%%%%%%%%%%%%%
\subsection{Some remarks}
Before starting the detailed
description of our method to  use the age measurements of a given
cosmic clock to get cosmological constraints, it worth considering
some caveats connected with our model. We concentrate in particular
on the Newtonian limit and  Post Parametrized Newtonian (PPN)
parameters constraints of this theory, and discuss some advantages
related to the use of the $t(z)$ relation, with respect to the
magnitude - redshift one, to constrain the cosmological parameters.
\subsubsection{Newtonian limit and  Post Parametrized
Newtonian (PPN) behaviour}

 Recently the cosmological relevance of extended  gravity
theories, as scalar tensor or high order theories, has been widely
explored. However, in the weak–limit approximation, all these
classes of theories should be expected to reproduce the Einstein
general relativity which, in any case, is experimentally tested only
in this limit. This fact is matter of debate since several
relativistic theories do not reproduce Einstein results at the
Newtonian approximation but, in some sense, generalize them, giving
rise, for example, to Yukawa – like corrections to the Newtonian
potential which could have interesting physical consequences.  In
this section we want to discuss the Newtonian limit of our model,
and study the Post Parametrized Newtonian (PPN) behaviour. In order
to recover the Newtonian limit, the metric tensor has to be decomposed as
 \beq\lb{metric}
 \gd=\eta\dmunu+h \dmunu\,,
 \eeq
 where $\eta\dmunu$ is the Minkoskwi metric and
$h\dmunu$ is a small correction to it. In the same way, we define
the scalar field $\psi$ as a perturbation, of the same order of the
components of $h\dmunu$, of the original field $\p$, that is
 \beq\lb{413}
 \p=\varphi_0+\psi\,,
 \eeq
where $\varphi_{0}$ is a constant of order unit. It is clear that
for $\varphi_{0}=1$ and $\psi=0$ Einstein general relativity is
recovered. To write in an appropriate form the Einstein tensor
$G\dmunu$, we define the auxiliary fields
 \beq\lb{metric2}
 \overline{h} \dmunu\equiv h\dmunu-\half\,\eta\dmunu h\,,
 \eeq
and
 \beq\lb{metric3}
 \sig\da\equiv {\overline h}_{\alp\beta,\gam}
 \eta^{\beta\gam}\,.
 \eeq
Given these definitions it turns out that the weak field limit of
the power-law potential gives (see \cite{nmc2})

 \begin{eqnarray}
 h_{00}&\simeq & \left[\varphi_0^{2}\frac{1-16
 \xi(s)}{2\xi(s)(1-12\,\xi(s))}\right]\frac{M}{r}\nonumber\\&-&\left[{M\,\varphi_0^{2}\over 1-12\xi(s)}\frac{\bar{V}_0 (p(s)-4)(p(s)-1)}{1-2\,\xi(s)
 }\right]r\\&-& 4\pi
 \left[\frac{ \bar{V}_0\varphi_0^{2+p(s)}}{\xi(s)}+\frac{ 2\varphi_0^{4}}{2 (p(s)-1)}\,\frac{\bar{V}_0 (p(s)-4)(p(s)-1)}{1-2\,\xi(s) }\right]r^2
  \label{428c} \nonumber\\
  h_{ij} & \simeq & \delta_{ij}\left\{
  \left[\varphi_0^{2}\frac{1-8 \xi(s)}{2\xi(s)(1-12\,\xi(s))}\right]\frac{M}{r}\right.\nonumber\\&+&\left[{M\,\varphi_0^{2}\over 1-12\xi(s)}\frac{\bar{V}_0 (p(s)-4)(p(s)-1)}{1-2\,\xi(s)
 }\right]r  \\
&+ & \left. 4\pi \left[\frac{
\bar{V}_0\varphi_0^{2+p(s)}}{\xi(s)}-\frac{ 2\varphi_0^{4}}{2
(p(s)-1)}\,\frac{\bar{V}_0 (p(s)-4)(p(s)-1)}{1-2\,\xi(s)
}\right]r^2\right\},\nonumber \label{ 428d}
 \end{eqnarray}
where only linear terms  in $V_0$ are given and we omitted the
constant terms.

In the case of the quartic potential case, instead we obtain:
\beqa
\lb{431}
 h_{00} & \simeq &\left[{1\over 2}\frac{\varphi_0^{2}(1-16\,\xi )}{\xi
 (1-12\,\xi )}\right]\frac{M}{r}-\left[\frac{4\pi
\lambda\varphi_0^{6}}{\xi}\right]r^2
 -\Theta\,,\\
 \lb{432}
 h_{il} & \simeq &
 \delta_{il}\left\{\left[{1\over 2}\frac{\varphi_0^{2}(1-8\xi)}{\xi
 (1-12\,\xi )}\right]\frac{M}{r}+
 \left[\frac{4\pi\lambda\varphi_0^{6}}{\xi}\right]r^2+
\Theta \right\} \,, \\
 \lb{433}
 \psi & \simeq & \left[{1\over 2}\frac{2}{(1-12\,\xi
)\varphi_0}\right]
 \frac{M}{r}+\psi_0,
 \eeqa
being $\Theta=2\varphi_0^{3}\psi_0 $ a sort of \textit{cosmological
term}, $\xi={3\over 32}$  and $\psi_0$ an arbitrary integration
constant. As we see, the role of the self--interaction potential is
essential to obtain corrections to the Newtonian potential, which
are  constant or quadratic as for other generalized theories of
gravity. Moreover, in general, any relativistic theory of gravitation can
yield corrections to the Newton potential (see for example
\cite{will}) which in the post-Newtonian (PPN) formalism, could
furnish tests for the same theory  based on local experiments. A
satisfactory description of PPN limit for scalar tensor theories has
been developed in (\cite{ damour2}). The starting point of such an
analysis consist in redefine the non minimally coupled Lagrangian
action in term of a minimally coupled scalar field model {\it via} a
conformal transformation from the Jordan to the Einstein frame. In
the Einstein frame deviations from General Relativity can be
characterized through Solar System experiments (\cite{will}) and
binary pulsar observations which give an experimental estimate of
the PPN parameters, introduced by Eddington to better determine the
eventual deviation from the standard prediction of General Relativity,
expanding the local metric as the Schwarzschild one, to higher order
terms. The generalization of this quantities to scalar-tensor
theories allows the PPN-parameters to be expressed in term of the
non-minimal coupling function $F(\phi)$ :
\begin{equation}\label{gamma}
\gamma^{PPN}-1\,=\,-\frac{F'(\phi)^2}{F(\phi)+2[F'(\phi)]^2},
\end{equation}
\begin{equation}\label{beta}
\beta^{PPN}-1\,=\,\frac{F(\phi){\cdot}
F'(\phi)}{F(\phi)+3[F'(\phi)]^2}\frac{d\gamma^{PPN}}{d\phi}.
\end{equation}
 Results about PPN parameters are summarized in Tab.\ref{ppn}.
\begin{table}
\centering
\begin{tabular}{|l|c|}
\hline\hline
  Mercury Perih. Shift&$|2\gamma_{0}^{PPN}-\beta_{0}^{PPN}-1|<3{\times}10^{-3}$ \\\hline
 Lunar Laser Rang. &  $4\beta_{0}^{PPN}-\gamma_{0}^{PPN}-3\,=\,-(0.7\pm 1){\times}{10^{-3}}$ \\\hline
 Very Long Bas. Int. &  $|\gamma_{0}^{PPN}-1|\,=\,4{\times}10^{-4}$ \\\hline
 Cassini spacecraft &  $\gamma_{0}^{PPN}-1\,=\,(2.1\pm 2.3){\times}10^{-5}$ \\\hline
\end{tabular}
\caption{\small \label{ppn} A schematic resume of recent constraints
on the PPN-parameters from Solar System experiments.}
\end{table}
The experimental results can be substantially resumed into the two
limits (\cite{schimd05})\,:
\begin{equation}\label{gblimits}
|\gamma^{PPN}_0-1|\leq{2{\times}10^{-3}}\,,\ \ \ \ \
|\beta_0^{PPN}-1|\leq{6{\times}{10^{-4}}}.
\end{equation} If we apply the formulae in the Eqs. (\ref{gamma}) and (\ref{beta})
to the power law potential, we obtain\,:
\begin{equation}\label{gamma2}
\gamma^{PPN}-1\,-\,\frac{4\xi(s)}{1+8\xi(s)}\,,
\end{equation}
\begin{equation}
\beta^{PPN}-1\,=\,\frac{F(\phi){\cdot}
F'(\phi)}{F(\phi)+3[F'(\phi)]^2}\frac{d\gamma^{PPN}}{d\phi}\,=0\,.
\end{equation}
%%\begin{equation}\label{beta}
%%=\,\frac{1}{2}\frac{\alpha^2}{(1+\alpha^2)^2}\frac{d\alpha}{d\phi}\,.
%%\end{equation}
The above definitions imply that the PPN-parameters in general
depend on the non-minimal coupling function $F(\phi)$ and its
derivatives. However in our model $\gamma^{PPN}$ depends only on $s$
while $\beta^{PPN}=\,1$.It turned out that the limits for
$\beta_0^{PPN}$ in the Eq. (\ref{gblimits}) is naturally verified,
for each value of $s$, while the constrain on $|\gamma^{PPN}_0-1|$
is satisfied for $s\in (-1.5,-1.4 )$ (see \cite{nmc2}). The case of
the quartic potential is quite different: we easily realize that the
constrain on $\gamma_0^{PPN}$ is not satisfied, while it is
automatically satisfied the constrain on $\beta_0^{PPN}$, as well
the  further limit on the two PPN-parameters $\gamma_0^{PPN}$ and
$\beta_0^{PPN}$, which can be outlined by means of the ratio (\cite{
damour2}):
\begin{equation}\label{pulsar}
\frac{\beta_0^{PPN}-1}{\gamma_0^{PPN}-1}\,<\,1.1\,.
\end{equation}
However the whole procedure of testing extended theories of gravity
based on local experiments (as PPN parameter constraints) subtends
two crucial questions of  theoretical nature: the question of the so
called  \textit{inhomogeneous gravity}, and the question of the
conformal transformations.
 The former mainly consist in matching local scales with the
cosmological background: there is actually no reason {\textit a
priori} why local experiments should match behaviours occurring
at cosmological scales. A non-minimally coupling function $F(\phi)$, for instance, can alter the
Hubble length at equivalence epoch, which is a scale
imprinted on the power spectrum. CMBR, large-scale structure and, in
general, cosmological experiments could provide complementary
constraints on extended theories. In this sense also the coupling
parameter $\xi \propto {1\over \omega_{BD}}$ may be larger than
locally is (\cite{CMB, francesca3}). It is actually very
interesting to compare just limits on the Brans-Dicke parameter $
\omega_{BD}$ coming both from  solar system experiments,
$\omega_{BD}>40000$ (\cite{will}), and from current cosmological
observations, including cosmic microwave anisotropy data and the
galaxy power spectrum data, $\omega_{BD}>120$)
(\cite{francesca3}).  As further consideration, it is worth
stressing that our  model $F(\phi)\propto (\phi+\phi_0)^2$ has
been derived  by requiring the existence of  Noether symmetries
for the pointlike Lagrangian defined in the FRW minisuperspace
variables $\{a,\dot{a},\phi,\dot{\phi}\}$: the extrapolation of
such a solution at local scales is, in principle, not valid since
the symmetries characterizing the cosmological model are not
working\footnote{In cosmology, we assume the Cosmological
Principle and then a dynamical behavior averaged on large scales.
This argument cannot be extrapolated to local scales, in
particular to Solar System scales, since anisotropies and
inhomogeneities cannot be neglected}. However, these
considerations do not imply that local experiments lose their
validity in probing alternative theories of  gravity \cite{will},
but point out the urgency to understand how  matching
observational results at different scales in a coherent and
self-consistent theory not available yet.
Also the question related to the conformal transformations is of great interest: these
ones are often used to reduce non minimally coupled scalar field
models to the cases of minimally coupled field models gaining wide
mathematical simplifications. The \textit{ Jordan frame}, in which
the scalar field is nonminimally coupled to the Ricci curvature,
is then mapped into the \textit{ Einstein frame} in which the
transformed scalar field is minimally coupled to the Ricci
curvature, but where appears an interaction term between standard
matter and the scalar field. It turns out the two frames are not
physically equivalent, and some care have to be taken in applying
such techniques (see for instance (\cite{faraoni2}) for a critical
discussion about this point). Actually even if a scalar tensor
theory of gravity in the Jordan frame can be mimicked by an
interaction between dark matter and dark energy in the Einstein
frame, the influence of gravity is quite different in each of them.  These considerations, however, do not imply that
local experiments lose their validity to probe scalar tensor theories
of gravity, but should simply highlight the necessity to compare
local and cosmological results, in order to understand how to
match different scales.
\subsubsection{Comparing between the $t(z)$  and the $(m-M)(z)$ relations}
It is well known that in scalar tensor theories of gravity, as well
in general relativity, the expansion history of the universe is driven
by the function $H(z)$; this implies that observational quantities,
like the luminosity distance, the angular diameter distance, and the
lookback time, are all function of $H(z)$ (in particular  $H(a)$
actually appears in the kernel of ones integral relations). It turns out
that the most appropriate mathematical tool to study the sensitivity
to the cosmological model of such observables  consists in
performing the functional derivative with respect to the
cosmological parameters (see \cite{saini} for a discussion about
this point in relation to distance measurements). In this section
however we face the question from an empirical point of view: we
actually limit to show that the lookback time is much more sensitive
to the cosmological model than other observables, like luminosity
distances, and the modulus of distance. This circumstance encourages
to use, together with other more standard techniques (as for
instance the Hubble diagram from SneIa observations), the age of
\textit{cosmic clocks} to test cosmological scenarios,
which could justify and explain the accelerated expansion of the
universe.

  In any case, before describing the lookback method in details,
let us briefly sketch some distance-based methods in order to
compare the two approaches.  It is well known that the use of
astrophysical standard candles provides a fundamental mean of
measuring the cosmological parameters. Type Ia supernovae  are the
best candidates for this aim since they can be accurately
calibrated and can be detected up to enough high red-shift. This
fact allows to discriminate among cosmological models. To this
aim, one can fit a given model to the observed
magnitude\,-\,redshift relation, conveniently expressed as\,:

\begin{equation}
\mu(z) = 5 \log{\frac{c}{H_0} d_L(z)} + 25 \label{eq: muz}
\end{equation}
being $\mu$ the distance modulus and $d_L(z)$ the dimensionless
luminosity distance.  Thus $d_L(z)$ is simply given as\,:

\begin{equation}
d_L(z) = (1 + z) \int_{0}^{z}{dz' [ \Omega_m (1 + z')^3 +
\Omega_{\Lambda} ]^{-1/2}} \ . \label{eq: dl}
\end{equation}
where $\Omega_ {\Lambda} = 1 - \Omega_m$ in the case of
$\Lambda$CDM model, but, in general,  $\Omega_ {\Lambda}$ can
represent any dark energy density parameter.

The distance modulus  can be obtained from observations of
SNe\,Ia. The apparent magnitude $m$ is indeed measured, while the
absolute magnitude $M$ may be deduced from template fitting or
using the Multi\,-\,Color Lightcurve Shape (MLCS) method. The
distance modulus is then simply $\mu = M - m$. Finally, the
redshift $z$ of the supernova can be determined accurately from
the host galaxy spectrum or (with a larger uncertainty) from the
supernova spectrum.

Roughly speaking, a given model can be fully characterized by two
parameters\,: the today Hubble constant $H_0$ and the matter
density $\Omega_m$.  Their best fit values can be obtained by
minimizing the $\chi^2$ defined as\,:

\begin{equation}
\chi^2(H_0, \Omega_m) = \sum_{i}{\frac{[\mu_i^{theor}(z_i | H_0,
\Omega_m) - \mu_i^{obs}]^2} {\sigma_ {\mu_{0},i}^{2} +
\sigma_{mz,i}^{2}}} \label{eq: defchi}
\end{equation}
where the sum is over the data points. In Eq.(\ref{eq: defchi}),
$\sigma_{\mu_{0}}$ is the estimated error of the distance modulus
and $\sigma_{mz}$ is the dispersion in the distance modulus due to
the uncertainty $\sigma_z$ on the SN redshift. We have\,:

\begin{equation}
\sigma_{mz} = \frac{5}{\ln 10} \left ( \frac{1}{d_L}
\frac{\partial d_L}{\partial z} \right ) \sigma_z \label{eq:
sigmamz}
\end{equation}
where we can assume $\sigma_z = 200 \ \rm{km \ s^{-1}}$ adding in
quadrature $2500 \ \rm{km \ s^{-1}}$ for those SNe whose redshift
is determined from broad features. Note that $\sigma_{mz}$ depends
on the cosmological parameters so that  an iterative procedure to
find the best fit values can be assumed.

For example, the High\,-\,z team and the Supernova Cosmology
Project have detected a quite large sample of high redshift ($z
\simeq 0.18 - 0.83$) SNe\,Ia, while the Calan\,-\,Tololo survey
has investigated the nearby sources. Using the data in Perlmutter
et al. \cite{Perl97} and Riess et al. \cite{Riess00}, combined
samples of  SNe  can be compiled giving confidence regions in the
$(\Omega_M, H_0)$ plane.

Besides the above method,   the Hubble constant $H_0$ and the
parameter $\Omega_{\Lambda}$ can be constrained also by the
angular diameter distance $D_A$ as measured using the
Sunyaev-Zeldovich effect (SZE) and the thermal bremsstrahlung
(X-ray brightness data) for galaxy clusters. Distances
measurements using SZE and X-ray emission from the intracluster
medium are based on the fact that these processes depend on
different combinations of some parameter of the clusters. The SZE
is a result of the inverse Compton scattering of the CMB photons
of hot electrons of the intracluster gas. The photon number is
preserved, but photons gain energy and thus   a decrement of the
temperature is generated in the Rayleigh-Jeans part of the
black-body spectrum while an increment rises up in the Wien
region. The analysis can be limited to the so called {\it thermal}
or {\it static} SZE, which is present in all the clusters,
neglecting the ${\it kinematic}$ effect, which is present in those
clusters with a nonzero peculiar velocity with respect to the
Hubble flow along the line of sight. Typically, the thermal SZE is
an order of magnitude larger than the kinematic one.
 The shift of temperature is:
\begin{equation}
\frac{\Delta T}{T_0} = y\left[ x \, \mbox{coth}\,
\left(\frac{x}{2} \right) -4 \right], \label{eq:sze5}
\end{equation}
where ${\displaystyle x=\frac{h \nu}{k_B T}}$ is a dimensionless
variable, $T$ is the radiation temperature, and $y$ is the so
called Compton parameter, defined as the optical depth $\tau =
\sigma_T \int n_e dl$ times the energy gain per scattering:
\begin{equation}\label{compt}
  y=\int  \frac{K_B T_e}{m_e c^2} n_e \sigma_T dl.
\end{equation}
In the Eq.~(\ref{compt}), $T_e$ is the temperature of the
electrons in the intracluster gas, $m_e$ is the electron mass,
$n_e$ is the numerical density of the electrons, and $\sigma_T$ is
the  cross section of Thompson electron scattering. The condition
$T_e \gg T$ can be adopted ($T_e$ is the order of $10^7\, K$ and
$T$, which is the CBR temperature is $\simeq 2.7K$). Considering
the low frequency regime of the Rayleigh-Jeans approximation one
obtains
\begin{equation}
 \frac{\Delta T_{RJ}}{T_0}\simeq -2y
 \label{eq:sze5bis}
\end{equation}
The next step  to quantify the SZE decrement is  to specify the
models for the intracluster electron density and temperature
distribution. The most commonly used model is the so called
isothermal $\beta$ model. One has
\begin{eqnarray}
& & n_e (r) = n_e (r) = n_{e_0} \left( 1 + \left(
\frac{r}{r_e} \right)^2 \right)^{-\frac{3 \beta}{2}} \\
& & T_e (r) = T_{e_0}~, \label{eq:sze6}
\end{eqnarray}
being $n_{e_0}$ and $T_{e_0}$, respectively the central electron
number density and temperature of the intracluster electron gas,
$r_e$ and $\beta$ are fitting parameters connected with the model.
Then we have
\begin{equation}
\frac{\Delta T}{T_0} = -\frac{2 K_B \sigma_T T_{e_0} \,
n_{e_0}}{m_e c^2} \cdot \Sigma \label{eq:sze7}
\end{equation}
being
\begin{equation}
\Sigma = \int^\infty_0 \left( 1 + \left( \frac{r}{r_c}
\right)^2\right)^{-\frac{3 \beta}{2}} dr\,. \label{eq:sze8}
\end{equation}
The integral in Eq.~(\ref{eq:sze8}) is overestimated  since
clusters have a finite radius.

A simple geometrical argument converts the integral in
Eq.~(\ref{eq:sze8}) in angular form, by introducing the angular
diameter distance, so that

\begin{equation}
\Sigma = \theta_c \left(1 + \left(
\frac{\theta}{\theta_2}\right)^2 \right)^{1/2 - 3 \beta/2}
\sqrt{\pi} \, \frac{\Gamma \left( \frac{3 \beta}{2} -
\frac{1}{2}\right)}{\Gamma \left( \frac{3 \beta}{2} \right)} \,
r_{DR}. \label{eq:sze9}
\end{equation}

In terms of the dimensionless angular diameter distances, $d_A$
(such that $D_A=\displaystyle\frac{c}{H_0} d_A$), one gets
\begin{equation}
\frac{\Delta T (\theta)}{T_0} = - \frac{2}{H_0}\frac{\sigma_T K_B
T_{ec}n_{e_0}}{m_e c} \sqrt{\pi}  \frac{\Gamma \left( \frac{3
\beta}{2} \frac{1}{2} \right)}{\Gamma \left(
\frac{3\beta}{2}\right)} \left( 1 - \left(
\frac{\theta}{\theta_2}\right)^2 \right)^{\frac{1}{2} (1 -3
\beta)} d_A, \label{eq:sze10}
\end{equation}
and, consequently, for the central temperature decrement, we
obtain
\begin{equation}\label{eq:sze10bis}
\frac{\Delta T (\theta =0)}{T_0}=- \frac{2}{H_0}\frac{\sigma_T K_B
T_{ec}n_{e_0}}{m_e c} \sqrt{\pi} \frac{\Gamma \left( \frac{3
\beta}{2} \frac{1}{2} \right)}{\Gamma \left(
\frac{3\beta}{2}\right)}\frac{c}{H_0} d_A.
\end{equation}
The factor $\displaystyle \frac{c}{H_0} d_A$  in
Eq.~(\ref{eq:sze10bis}) carries the dependence on the thermal SZE
on both the cosmological models (through $H_0$ and the Dyer-Roeder
distance $d_A$) and the redshift (through $d_A$). From
Eq.~(\ref{eq:sze10bis}), we also note that the central electron
number density is proportional to the inverse of the angular
diameter distance, when calculated through SZE measurements. This
circumstance allows to determine the distance of cluster, and then
the Hubble constant, by the measurements of its thermal SZE and
its X-ray emission.

This possibility is based on the different power laws, according
to which  the decrement of the temperature in the SZE,
$\frac{\Delta T(\theta =0)}{T_0}$ , and the X-ray emissivity,
$S_X$, scale with respect to the electron density. In fact, as
pointed out, the electron density, when calculated from SZE data,
scales as $d^{-1}_{A}$~( $n^{SZE}_{e0}\propto d^{-1}_A$), while
the same one scales as $d^{-2}_{A}$~($n^{X-ray}_{e0}\propto
d^{-2}_A$) when calculated from X-ray data. Actually, for the
X-ray surface brightness, $S_X$,  assuming for the temperature
distribution of  $T_e=T_{e0}$, one gets the following formula:
\begin{equation}\label{sx}
 S_X=\frac{\epsilon_X}{4\pi}{n_{e0}^2}\frac{1}{ (1+z)^3} \theta_c \frac{c}{H_0} d_A I_{SX},
\end{equation}
being \[I_{Sx}=\int^{\infty}_0 \left(\frac{n_e}{n_{e0}}\right)^2
dl\] the X-ray structure integral, and $\epsilon_X$ the spectral
emissivity of the gas (which, for $T_e\geq 3{\times}10^{7}$, can
be approximated by a typical value: $\epsilon_X = \epsilon
\sqrt{T_e}$, , with $\epsilon \simeq 3.0 {\times} 10^{-27} n_p^2$
erg $cm^{-3}$ $s^{-1}$ $K^{-1}$). The angular diameter distance
can be deduced by eliminating the electron density from
Eqs.~(\ref{eq:sze10bis}) and (\ref{sx}), yielding:

\begin{equation}
\frac{y^2}{S_X}= \frac{4 \pi (1+z)^3}{\epsilon} {\times}
 \left(\displaystyle\frac{k_B \sigma_{T}}{m_e c^2}\right)^2 {T_{e0}}^{3/2}
 \theta_c \frac{c}{H_0} d_A {\times}
\frac{\left[B(\frac{3}{2}\beta-\frac{1}{2},
\frac{1}{2})\right]^2}{B(3\beta-\frac{1}{2}, \frac{1}{2})}
\label{eq:sze11}
\end{equation}
where $B(a,b)=\displaystyle\frac{\Gamma(a)\Gamma(b)}{\Gamma(a+b)}$
is the Beta function.

It turns out that
\begin{equation}\label{eq:sze11bis}
D_A=\frac{c}{H_0}d_A\propto \frac{(\Delta T_0)^2}{S_{X0}
T^2_{e0}}\frac{1}{\theta_c},
\end{equation}
where all these quantities are evaluated along the line of sight
towards the center of the cluster (subscript 0), and $\theta_c$ is
referred to a characteristic scale of the cluster along the line
of sight. It is evident that the specific meaning of this  scale
depends on the density model adopted for  clusters. In general,
the so called $\beta$ model is used.

Eqs.~(\ref{eq:sze11}) allows  to  compute the Hubble constant
$H_0$, once the redshift of the cluster is known and the other
cosmological parameters are, in same way, constrained. Since the
dimensionless Dyer-Roeder distance, $d_A$,  depends on
$\Omega_{{\Lambda}}$, $\Omega_{m}$, comparing the  estimated
values with the theoretical formulas for $D_A$, it is possible to
obtain information about $\Omega_m$, $\Omega_{{\Lambda}}$, and
$H_0$. Modelling the intracluster gas as a spherical isothermal
$\beta $-model allows to obtain constraints on the Hubble constant
$H_0$ in a standard $\Lambda$CDM model. In general, the results
are in good agreement with those  derived from SNe Ia data.

\begin{figure}[ptb]
\centering{ \includegraphics[width=7 cm, height=6.5
cm]{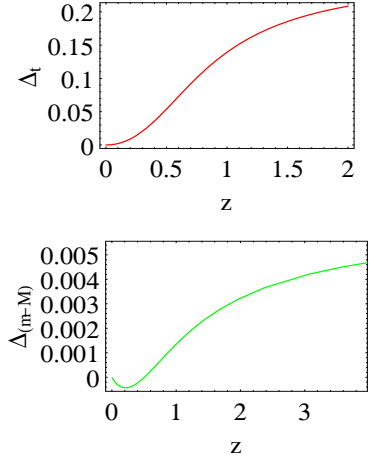}\caption{\small We compare the sensitivity to the
values of the parameters in our quartic potential model in the
lookback time relation and in the modulus of distance. Actually  we
plot the relative variation in $t(z)$ (upper diagram) and $m-M$ with
respect to a variation of $\alpha_1$ from $3.$ to $ 3.5$ (the other
parameters are fixed). It turns out that the $t(z)$ relation is much
more sensible. }} \label{compare}
\end{figure}

\begin{figure}
\centering{ \includegraphics[width=5 cm, height=5
cm]{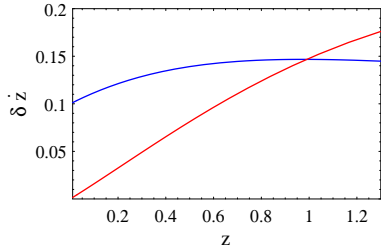}}\caption{\small We compare the sensitivity of the
${dz \over dt}$ relation to the values of the parameters in the
power law potential cosmological model. Actually  we plot the
relative variation in ${dz \over dt}$ with respect to a variation of
$s$ from $-1.4$ to $ -1.3$ (red line), and with respect to a
variation of $\widehat{H_0}$ from $1$ to\,$.9$ (blue line), the
other parameters being fixed. }\label{comparez}
\end{figure}

Apart from the advantage to provide an further instrument of
investigation, the lookback time method uses the stronger
sensitivity to the cosmological model, characterizing  the $t(z)$
relation, as shown in Figs. (\ref{compare},\ref{comparez}).
Moreover, as we will discuss in the next sections such a method
reveals its full validity when applied to old objects at very high
$z$: actually it turns out that this kind of analysis is very
strict and could remove, or at least reduce, the degeneracy which
we observe at lower redshifts, also, for instance, considering the
Hubble diagram observations, where  different cosmological models
allow to fit with the same statistical significance several
observational data. In a certain sense, we could argue that the
lookback time, even if exhibits a wide mathematical  homogeneity
with the distance observables, does not contain the same
information, but, rather, presents some interesting peculiar
properties, useful to investigate also alternative gravity
theories. However it is important to remark that the comparison
with observational data in scalar tensor theories of gravity is
more complex than in general relativity, since the action of
gravity is different. For instance, the use of type Ia supernovae
to constraint the cosmological parameters (and hence the claim
that our universe is accelerating) mostly lies in the fact that we
believe that they are standard candles so that we can reconstruct
the luminosity distance vs redshift relation and compare it with
its theoretical value. In a scalar-tensor theory, we have to
address both questions, i.e: the determination of the luminosity
distance vs redshift relation  and the property of standard candle
since two supernovae at different redshift probe different
gravitational coupling constant and could not be standard candles
anymore. Actually it can be shown that in this case it is possible
to \textit{generalize} the theoretical expression of the distance
modulus, taking into account the effect of the variation of
$G_{eff}$ through a correction term (\cite{Gat01}). Of course, the
variation of $G_{eff}$ with time (and then redshift) could
challenges the reliability of age measurements of some cosmic
clocks to test cosmological models, unless to construct some
reasonable theoretical model, which quantifies and corrects this
effect. Actually the estimation of cluster ages is based on
stellar population synthesis models which rely on  stellar
evolution models formulated in a Newtonian framework. In order to
be consistent with scalar tensor gravity, which assume  an
evolving gravitational coupling $G_{eff}$, one should investigate
to which extent this $G_{eff}$ variation affects the results of
the given stellar population synthesis model. However, it turns
out that in the context of our qualitative analysis, the effective
gravitational coupling varies in a range of no more than $6\%$ and
then the effect of such a variation  can be included in a bias
factor $df$, which, resulting to be $df=4.5\pm 0.5$, as we will
see in the next section,  affects the age measurements of the
considered cluster sample much more than the variation of
$G_{eff}$.

%%%%%%%%%%%%%%%%%%%%%%%%%%%%%%%%%%%%
\section{The dataset at low and intermediate redshifts}
\begin{figure}[ptb] \centering{ \includegraphics[width=7 cm,
height=6.5 cm]{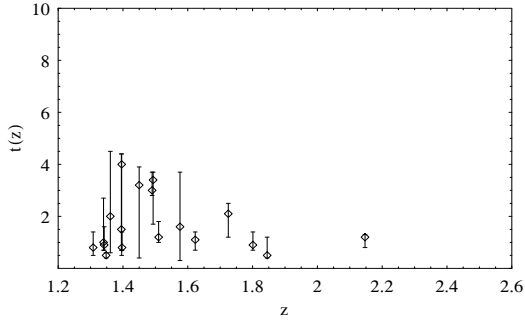}} \caption{{\small Original data from
GDDS. This sample corresponds to 20 old passive galaxies distributed
over the redshift interval $1.308 <z< 2.147$, as given by McCarthy
et al. (2004).}} \label{newdata2}
\end{figure}

\begin{figure}[ptb] \centering{ \includegraphics[width=7 cm,
height=6.5 cm]{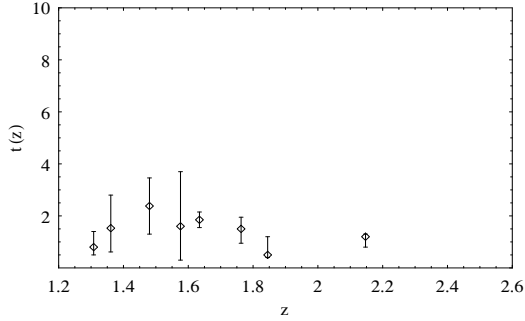}} \caption{{\small The 8 high-z
measurements selected after the criterion discussed in the text.}}
\label{newdata}
\end{figure}

%%%%%%%%%%%%%%%%%%%%%%%%%%%%%%%%%%%%
In order to discuss age constraints for the  above background
models, we first use the dataset compiled by
Capozziello {\it et al.} (\cite{look}), and given in Table 1, which
consists of age estimates of galaxy clusters at six redshifts
distributed in the interval $0.10\leq z \leq 1.27$ (see Sec. IV of (
\cite{look}) for more details on this age sample). To extend such a
dataset to higher redshifts, we join the GDDS sample presented by
McCarthy et al. (2004), consisting of 20 old passive galaxies,
distributed over the redshift interval $1.308 \leq z \leq 2.147$,
and  shown in Fig. (\ref{newdata2}). In order to build up our total
loockback time  sample, we first select from GDDS observations what
we will consider as the most appropriate data to our cosmological
analysis. Following (Dantas et al.\cite{dantas}) we adopt the
criterion that given two objects at (approximately) the same $z$,
the oldest one is always selected, ending up with a sample of 8 data
points, as in Fig.(\ref{newdata}).

\begin{table}[ptb]
\begin{center}%
\begin{tabular}
[c]{c|c|cc|c|c|c|c}\hline \multicolumn{4}{c|}{Color age} &
\multicolumn{4}{|c}{Scatter age}\\\hline $z$ & N & Age (Gyr) &  &
$z$ & N & Age (Gyr) & \\\hline\hline
0.60 & 1 & 4.53 &  & 0.10 & 55 & 10.65 & \\
0.70 & 3 & 3.93 &  & 0.25 & 103 & 8.89 & \\
0.80 & 2 & 3.41 &  & 1.27 & 1 & 1.60 & \\\hline
\end{tabular}
\end{center}
\caption{Main properties of the cluster sample compiled by
Capozziello {\it et al.} (\cite{look}) used for the analysis. The
data in the left part of the Table refers to clusters whose age has
been estimated from the color of the reddest galaxies (color age),
while that of clusters in the right part has been obtained by the
color scatter (scatter age). For each data point, we give the
redshift $z$, the number $N$ of clusters used and the age estimate}
\end{table}

Through the Eqs. (\ref{eq: titl}) and (\ref{eq: deftlobs}) in
Appendix A, which in our case can only be evaluated numerically, we
perform a $\chi^{2}$ analysis. For the power law potential we
obtain, in our units, $\chi_{red}^{2}=0.95$, $\mathcal{H}_{0}=1.00_{-0.07}%
^{+0.01}$, $s=-1.39_{-0.4}^{+0.5}$, $t_{0}=14.04\pm.08$ Gyr, and
$df=3.6\pm 0.7$ Gyr. Such best fit values correspond to $\Omega_{\Lambda_{eff}%
}=0.76_{-0.08}^{+0.03}$, according to the Eq. (\ref{omegaeff}).
\begin{figure}[ptb] \centering{ \includegraphics[width=7 cm,
height=6.5 cm]{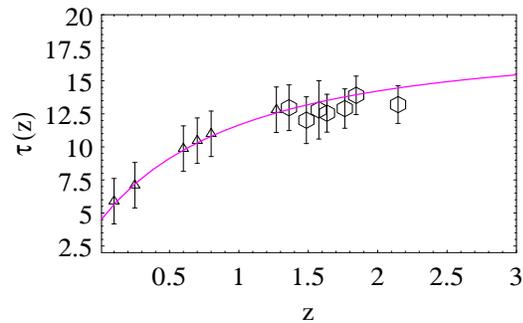}} \caption{{\small The observational
data of the whole dataset fitted to our model, with $\mathcal{H}_{0}=1.00_{-0.07}%
^{+0.01}$, $s=-1.39_{-0.4}^{+0.5}$, $t_{0}=14.04\pm.08$ Gyr, and
$df=3.6\pm 0.7$ Gyr. .}}
\label{bestfits}%
\end{figure}

\begin{figure}[ptb] \centering{ \includegraphics[width=7 cm,
height=6.5 cm]{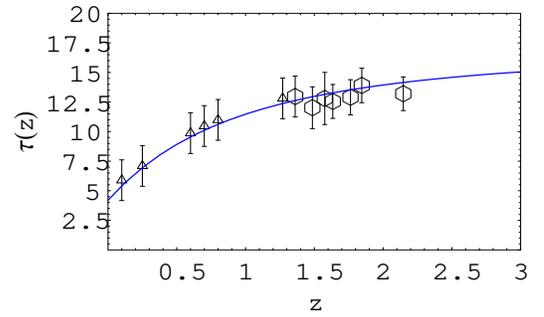}} \caption{{\small The observational
data of the whole dataset fitted to our quartic potential model, with $\mathcal{H}_{0}=1.00_{-0.05}%
^{+0.01}$, $\alpha_3=2.5_{-0.1}^{+0.5}$, $t_{0}=13.04\pm.05$ Gyr,
and $df=4.2\pm 0.7$ Gyr.}} \label{bestfitq}
\end{figure}

In the case of  quartic potential we obtain
$\chi_{red}^{2}=1.01$, $\mathcal{H}_{0}=1.00_{-0.05}%
^{+0.01}$, $\alpha_3=2.5_{-0.1}^{+0.5}$, $t_{0}=13.04\pm.05$ Gyr,
and $df=4.2\pm 0.7$ Gyr. In the Figs.(\ref{bestfits} and
\ref{bestfitq}), the observational data are plotted vs our best fit cosmological model.

\textit{Remark: }The range of values for $\mathcal{H}_{0}$ does not
correspond to a variation in the physical value of $H_{0}$, which is a prior for the model. It reflects instead the scatter in the universe age.

%\begin{figure}[ptb]
%\centering{ \includegraphics[width=7 cm, height=6.5
%cm]{comparew.eps}\caption{\small Analogously to the situation
%illustrated above, we compare the sensitivity to the values of the
%equation of state $w$ for a standard quintessence model with
%$w=constant $ in the lookback time relation and in the modulus of
%distance. Actually  we plot the relative variation in $t(z)$
%(upper diagram) and $m-M$ with respect to a variation of $w$ from
%$-1$ to $-0.23$ (of course $w$ increases from the bottom to the
%top in each diagram).  }} \label{comparew}
%\end{figure}

%\begin{figure}[ptb] \centering{ \includegraphics[width=7 cm,
%height=6.5 cm]{relativevar.eps}\caption{\small Relative variation of
%$G_{eff}$ with respect to the actual value, as function of $z$: it
%turns out that in our redshift range it is quite small, and does not
%exceed few $\%$. }} \label{relativevar}
%\end{figure}

\section{Extending the analysis at high redshifts}

%%%%%%%%%%%%%%%%%%%%%%%%%%%%%%%%%%%%%%%%%%%%%%%%%%%%%%%%%
Previous discussion shows that the scalar-tensor quintessence model
which we are studying can be successfully constrained by cosmic
clocks (clusters of galaxies) at low $(z\sim0\div0.5)$ and
intermediate $(z\sim1.0\div1.5)$ redshifts. In this section we
investigate its viability vs the age estimates of
some high redshift objects, with a minimal stellar age of 1.8
Gyr, 3.5 Gyr and 4.0 Gyr, respectively. It is actually well known
that the evolution  of the universe age with redshift (${\frac
{\displaystyle dt_{U}}{dz}}$) differs from a scenario to another;
this means that models in which the universe is \textit{old enough}
to explain the total expansion age at $z=0$ may not be compatible
with age estimates of high redshifts objects. This reinforces the
idea that dating of objects constitutes a powerful methods to
constrain the age of the universe at different stages of its
evolution (\cite{KC03,dun96,ferreras03}), and the first epoch of the
quasar formation can be a useful tool for discriminating among
different scenarios of dark energy (\cite{Jimenez2,al03}). The existence of some recently reported old
high-redshift objects if of relevance here, namely LBDS 53W091, a 3.5 Gyr-old radio
galaxy at $z=1.55$, LBDS 53W069, a 4 Gyr-old radio-galaxy at
$z=1.43$, and APM 08279+5255, an old quasar at $z=3.91$, whose age
was firstly estimated to lie in the range $2\div 3$ Gyr
(\cite{hasinger02}), and then updated to the range $1.8\div2.1$ Gyr
(\cite{fr05}). It is clear that these objects can be used to impose
more strict constraints on our model.

We take advantage for the fact that we have exact solutions, so that the redshift-time relation can be inverted. It is
easily derived from Eq.(\ref{scale}).

Taking for granted that the universe age at any redshift must
be greater than or at most equal to the age of the oldest object
contained in it, we introduces the the ratio
\[
\mathcal{R}\equiv{\frac{t_{z}}{t_{g}}}\geq1,
\]
as in (\cite{al03}), with $t_{z}$ derived from Eq. (\ref{scale}),
and $t_{g}$  the measured age of the objects. For each
extragalactic object, this inequality defines a dimensionless
parameter $T_{g}=H_{o}t_{g}$. In particular, for LBDS 53W091 radio
galaxy discovered by Dunlop et al.
(\cite{dun96}), the lower limit for the age yields $t_{g}(z=1.55)=3.5H_{0}%
\ $\textrm{{Gyr}} which takes values in the interval $0.21\leq
T_{g}\leq0.28$. It therefore follows that $T_{g}\geq0.21$. Similar
considerations may also be applied to the $4.0$ Gyr old galaxy at
$z=1.43$, for which $T_{g}\geq0.24$, and to the APM 08279+5255,
which corresponds to $T_{g}\geq0.131$.

Only models having an expanding age parameter larger than  $T_{g}$ at the above values of redshift will
be compatible with the existence of such objects.

\begin{figure}[ptb]
\centering{ \includegraphics[width=5 cm, height=7 cm]{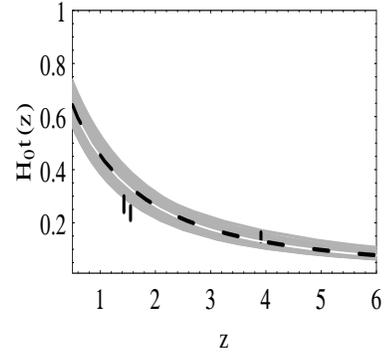}}
\caption{{\small The dashed line indicates $T(z)=H_{0}t(z)$
according to the best fit values of Sec. IV. for the power law
potential model. The shadowed region covers the allowed ranges
coming from lookback time tests. The vertical lines give the lowest
ages }$T_{g}${\small  of the considered objects for the allowed
range of }$h=(0.64\div 0.80)$. The figures reported in the text
correspond to $h=0.64$, i.e. the lowest edges of the lines.For the
first two objects, there is full agreement with the requirement
$T>T_{g}$. The third age is marginally compatible, likely  due to
systematics in evaluating the age-metalicity relation.}
\label{ages2}
\end{figure}

\begin{figure}[ptb]
\centering{ \includegraphics[width=5 cm, height=7 cm]{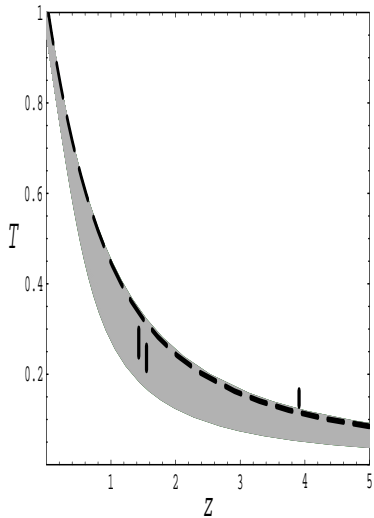}}
\caption{{\small The same than in Fig.(\ref{ages}) for the quartic
potentail model}} \label{ages}
\end{figure}

In Figs. (\ref{ages2}), and (\ref{ages}) we show the diagram of the
dimensionless age parameter $T(z)=H_{0}t(z)$ as a function of the
redshift  for two \textit{extreme} best fit values of the parameters
($s$ and $\mathcal{H}_{0}$ in the case of power law models, $\alpha_3$ and  $\mathcal{H}_{0}$ for the quartic potential model),
coming from the lookback time test. We observe that the age test
turns out to be quite critical, even though such values of the
parameters are fully compatible and \textit{equivalent} to the
$\chi^{2}$ analysis of other datasets (see \cite{nmc,nmc2}).

In fact, it turns out that the value $T_{g}\geq 0.131$ (\cite{fr05})
at $z=3.91$ is quite selective in both cases.

A remark is necessary at this point: the range of ages for quasar
APM 08279+5255 is $1.8$ and $2$ Gyr. It is required for models from
$10^{11}$ M$_{\odot}$ to $10^{12}$ M$_{\odot}$ in order to reach the
metalicity Fe/O=$2.5\times$ solar unit (for details see
\cite{fr05}).

%%%%%%%%%%%%%%%%%%%%%%%%%%%%%%%%%%%%%%%%%%%%%%%%%%

\section{Discussion and Conclusions}

%%%%%%%%%%%%%%%%%%%%%%%%%%%%%%%%%%%%%%%%%%%%%%%%%%
Cosmological models can be constrained not only using distance
indicators but also cosmic clocks, once efficient methods are
developed to estimate the age of distant objects. Among these, the
relations between lookback time and redshift $z$ are particularly
useful to discriminate among the huge class of dark energy models
which have been recently developed to explain the observed present
day accelerated behaviour of cosmic flow. In a near future, beside
distance measurements, time measurements could greatly contribute to
achieve a final cosmological model constraining the main
cosmological parameters.

In this context, it is of fundamental importance to obtain valid
cosmic clocks at low, intermediate and high redshifts in order to
fit, in principle, a model at every epoch. In this case, degeneracies
are removed and the reliability of the model can be proved.

In this paper, we have tested a non minimally coupled scalar-tensor
quintessence model, characterised by quartic self-interacting
potential, using, as cosmic clocks, cluster of galaxies at low and
intermediate redshifts, two old radio galaxies at high redshift, and
a very far quasar. The results are comfortable since the model seems
to work in all the regimes considered. However, to be completely
reliable, the dataset should be further enlarged and the method
considered also for other time indicators.

The main feature of this approach is the fact that cosmic clocks are
completely independent of each other, so that, in principle, it is
possible to avoid biases due to primary, secondary and so on
indicators, as in cosmic ladder method. In that case, every rung of
the ladder is affected by the errors of former ones and it affects
the successive ones. By using cosmic clocks, this shortcoming can be,
in principle, avoided, since indicators are, by definition,
independent. In fact, we have used different methods to test the
model at low and high redshift with different indicators, which seem
to confirm independently the proposed dark energy model.

Another comment is due at this point. Having normalized the Hubble
parameter at present epoch, assuming $\mathcal{H}_{0}=H(1)$, we do
not need, in principle, priors on such a parameter, since we are
using an exact solution. In other words, we can check the validity
of the model by selecting reliable cosmic clocks only. Another
advantage of such a choice is that one has to handle only small
numbers of parameters in numerical computations. \\

\section*{ Acknowledgements} This research was supported by the
National Research Foundation (South Africa) and the Italian {\it
Ministero Degli Affari Esteri-DG per la Promozione e Cooperazione
Culturale} under the joint Italy/ South Africa Science and
Technology agreement.

\appendix

\section{The lookback time method}

In order to use the age measurements of a given cosmic clock to
get cosmological constraints, let us consider an object $i$ at
redshift $z$ and let $t_{i}(z)$ be its age defined as the
difference between the age of the universe at the formation
redshift $z_{F}$, and at $z$:
\begin{equation}
t_{i}(z)=t_{L}(z_{F})-t_{L}(z). \label{eq: titl}%
\end{equation}
If one is able to to estimate the ages $\{t_{i}\}$ for
$i=1,2,\ldots,N$ of $N$ objects, we can estimate the lookback time
$t_{L}^{obs}(z_{i})$ as\thinspace\
\begin{eqnarray}
t_{L}^{obs}(z_{i})  &  =t_{L}(z_{F})-t_{i}(z)\nonumber\\
~  &  =t_{0}^{obs}-t_{i}(z)-df\ , \label{eq: deftlobs}%
\end{eqnarray}
where $t_{0}^{obs}$ is the age of the universe (which in our units
is set to
1), while the bias (\textit{delay factor}) can be defined as\thinspace:%

\begin{equation}
df=t_{0}^{obs}-t_{L}(z_{F})\ .
\end{equation}
The delay factor is introduced to take into account our ignorance
of the formation redshift $z_{F}$ of the object. Actually, to
estimate $z_{F}$, one should use Eq.(\ref{eq: titl}), assuming a
background cosmological model. Since our aim is to constrain the
background cosmological model, it is clear that we cannot infer
$z_{F}$ from the measured age, so that this quantity is \textit{a
priori} undetermined. Moreover we rely that it can take into
account also the effect of the $G_{eff}$ variation on the age
estimations, because the expected magnitude of such an effect. In
principle, $df$ should be different for each object in the sample
unless there is a theoretical reason to assume the same redshift
at the formation of all the objects. However, we can realistically
assume that $df$ is the same for all the homologous objects of a
given dataset (in the range of the errors), and we consider $df$
rather than $z_{F}$ as the unknown parameter to be determined from
the data.

We may then define a merit function \thinspace:
\[
\chi_{lt}^{2}=\displaystyle{\frac{1}{N-N_{p}+1}}\left\{  \left[
\frac
{t_{0}^{theor}(\mathbf{p})-t_{0}^{obs}}{\sigma_{t_{0}^{obs}}}\right]
^{2}+\sum_{i=1}^{N}{\left[  \frac{t_{L}^{theor}(z_{i},\mathbf{p})-t_{L}%
^{obs}(z_{i})}{\sqrt{\sigma_{i}^{2}+\sigma_{t}^{2}}}\right]
^{2}}\right\}  ,
\]
where $N_{p}$ is the number of parameters of our model,
$\sigma_{t}$, $\sigma_{i}$ are the uncertainties on $t_{0}^{obs}$
and $t_{L}^{obs}(z_{i})$. Here the superscript \textit{theor}
denotes the predicted values of a given quantity.

In principle, such a method can work efficiently to discriminate
between the various cosmological models. However, the main
difficulty is due to the lack of available data which leads to
large uncertainties on the estimated parameters. In order to
partially alleviate this problem, we can add further constraints
on the model by using \textit{priors}; for example choosing a
Gaussian prior on the Hubble constant allows us to redefining the
likelihood function as
\begin{equation}
{\mathcal{L}}(\mathbf{p})\propto{\mathcal{L}}_{lt}(\mathbf{p})\exp{\left[
-\frac{1}{2}\left(  \frac{h-h^{obs}}{\sigma_{h}}\right)  ^{2}\right]  }%
\propto\exp{[-\chi^{2}(\mathbf{p})/2],} \label{eq: deflike}%
\end{equation}
where we have absorbed $df$ to the set of parameters $\mathbf{p}$
and have defined\thinspace:
\begin{equation}
\chi^{2}=\chi_{lt}^{2}+\left(  \frac{h-h^{obs}}{\sigma_{h}}\right)
^{2}
\label{eq: newchi}%
\end{equation}
with $h^{obs}$ the estimated value of $h$ and $\sigma_{h}$ its
uncertainty. We use the HST Key project results (\cite{Freedman})
setting $(h,\sigma _{h})=(0.72,0.08)$. Note that this estimate is
independent of the cosmological model since it has been obtained
from local distance ladder methods.

The best fit model parameters $\mathbf{p}$ may be obtained by
maximizing ${\mathcal{L}}(\mathbf{p})$ which is equivalent to
minimize the $\chi^{2}$ defined in Eq.(\ref{eq: newchi}). It is
worth stressing that such a function should not be considered as a
\textit{statistical $\chi^{2}$} in the sense that it is not forced
to be of order 1 for the best fit model to be considered as a
successful fit. Actually, such an interpretation is not possible
since the errors on the measured quantities (both $t_{i}$ and
$t_{0}$) are not Gaussian distributed. Moreover, there are
uncontrolled systematic uncertainties that may also dominate the
error budget. Moreover, there are uncontrolled systematic
uncertainties that may also dominate the error budget.
Nevertheless, a qualitative comparison of different models may be
obtained by comparing the values of this pseudo $\chi^{2}$, even
if this should not be considered as definitive evidence against a
given model.

Given that we have more than one parameter, we obtain the best fit
value of each single parameter $p_{i}$ as the value which
maximizes the marginalized
likelihood for that parameter, defined as\thinspace:%

\begin{equation}
{\mathcal{L}}_{p_{i}} \propto\int{dp_{1} \ldots\int{dp_{i - 1}
\int{dp_{i + 1}
\ldots\int{dp_{n} \ {\mathcal{L}}(\mathbf{p})}}}} \ . \label{eq: deflikemar}%
\end{equation}
After having normalized the marginalized likelihood to 1 at maximum,
we compute the $1 \sigma$ and $2 \sigma$ confidence limits (CL) on
that parameter by solving ${\mathcal{L}}_{p_{i}} = \exp{(-0.5)}$ and
${\mathcal{L}}_{p_{i}} = \exp{(-2)}$ respectively.

\end{document}